\documentclass[final,5p,times,twocolumn]{elsarticle}
\usepackage{graphicx}
\usepackage{txfonts}
\usepackage[british]{babel}
\usepackage{latexsym}
\usepackage{amssymb}
\journal{New Astronomy}

\begin{document}

\begin{frontmatter}



\title{On turbulent fragmentation and the origin of the stellar IMF}


\author[SVA]{Anathpindika, S.}
\ead{sumedh$\_$a@iiap.res.in}
\address{Indian Institute of Astrophysics, 2$^{nd}$-Block, Koramangala, Bangalore-560034, India}

\begin{abstract}

 Two varieties of the universal stellar initial mass function (IMF) viz., the Kroupa and the Chabrier IMF, have emerged over the last decade to explain the observed distribution of stellar masses. The possibility of the universal nature of the stellar IMF leads us to the interesting prospect of a universal mode of star-formation. It is well-known that turbulent fragmentation of gas in the interstellar medium produces a lognormal distribution of density which is further reflected by the mass-function for clumps at low and intermediate masses. Stars condense out of unstable clumps through a complex interplay between a number of dynamic processes which must be accounted for when tracing the origin of the stellar IMF. In the present work, applying the theory of gravitational fragmentation we first derive the mass function (MF) for clumps. Then a core mass function (CMF) is derived by allowing the clumps to fragment, having subjected each one to a random choice of gas temperature. Finally, the stellar IMF is derived by applying a random core-to-star conversion efficiency, $\epsilon$, in the range of 5\%-15\% to each CMF. We obtain a power-law IMF that has exponents within the error-bars on the Kropua IMF. This derived IMF is preceded by a similar core mass function which suggests, gravoturbulent fragmentation plays a key role in assembling necessary conditions that relate the two mass-functions. In this sense the star-formation process, at least at low redshifts where gas cooling is efficient, is likely to be universal. We argue that the observed knee in the CMF and the stellar IMF may alternatively be interpreted in terms of the characteristic temperature at which gas in potential star-forming clouds is likely to be found. Our results also show that turbulence in star-forming clouds is probably driven on large spatial scales with a power-spectrum steeper than Kolmogorov-type.
\end{abstract}

\begin{keyword}
Star-formation -- gravoturbulent fragmentation -- Mass functions
\end{keyword}
\end{frontmatter}

\section[]{Introduction}
Explaining the origin of the stellar initial mass function (IMF) is an outstanding question before the theory of star-formation. Ever since the seminal work by Edwin Salpeter (Salpeter 1955) on this subject, first suggesting a power-law mass-distribution of stellar masses for stars more massive than $\sim$1 M$_{\odot}$, astronomers have toiled hard to explain its origin. Early work on the subject attempted to explain it by invoking the possibility of successive fragmentation of unstable clumps (Larson 1973). Though it was later shown that such fragmentation cannot continue  unbridled and must be arrested at some point; the minimum mass of a fragment was calculated by Low \& Lynden-Bell (1976). One of the first propositions for a universal IMF after Larson (1973), was made by Miller \& Scalo (1979), who having studied a number of nearby star-forming regions proposed a lognormal distribution. The question has been receiving greater attention in recent years, especially in view of the reported similarities between the distribution of core masses and the stellar IMF; see for example, Motte \emph{et al.} 1998, Nutter \& Ward-Thompson 2007, and Simpson \emph{et al.} 2011, amongst a number of other authors.

 While molecular clouds(MCs) contract under self-gravity, the presence of a turbulent velocity field fragments it into a number of clumps and filaments. With growing evidence pointing towards the turbulent nature of MCs, a number of authors have attempted to simulate the effect of turbulence. Numerical models  by for instance, V{\`a}zquez-Semadeni (1994), and Padoan \& Nordlund (1999) showed that turbulence generates a lognormal density distribution in the volume of gas. Though turbulence is dissipative and decays relatively quickly, usually on a timescale comparable to the sound-crossing time, it delays the onset of gravitational instability (see for e.g., Scalo \& Pumphrey 1982, MacLow \emph{et al.} 1998, Stone \emph{et al.} 1998, and Klessen \emph{et al.} 2000).  Some of the more massive clumps once dominated by self-gravity, begin to collapse, where-as other smaller clumps are more likely to be sheared apart by turbulent flows.

According to the fragmentation model introduced by Padoan \& Nordlund (2002; PN), the mass distribution of clumps forming out of a turbulent density field reflects the distribution of gas density. In recent years two variants of this theory have been introduced. These theories have a subtle variation in the density threshold at which a volume of gas is likely to support fragmentation. In the PN theory this threshold is fixed by the thickness of the shocked layer resulting from collisions between turbulent flows. On the other hand, in the Krumholz \& McKee (2005; KM) theory, the density threshold is set by equating the fastest growing unstable mode with the classical Jeans length. Finally, in the Hennebelle \& Chabrier (2008; HC) theory, there is no density threshold but instead, density perturbations are assumed to condense out as unstable clumps.

 Numerical simulations by Padoan \& Nordlund (2007) have demonstrated the power-law nature of clumps resulting from turbulent fragmentation. More recent simulations such as those by Federrath \emph{et al.}(2010), argue about the possible effects of the modes of driven turbulence, compressive against solenoidal. The resulting density distribution according to these latter authors depends on the driven mode. A possible short-coming of the theory of turbulent fragmentation is its failure to predict beyond the mass-spectrum for gas clumps, and therefore the stellar IMF itself, since not all clumps actually spawn stars. While sophisticated numerical algorithms can possibly make the simulations more realistic by including effects of energy feedback, tracing the evolution of gas over several orders of magnitude, up to stellar densities, exacts enormous computational costs. And so, a semi-analytic scheme like the one presented here can shed some light on the problem. 

Veltchev \emph{et al.}(2011) for example, recently presented one such calculation where the mass function for unstable clumps, cores, and stars was obtained. Having derived the mass function for clumps and cores using respectively, the turbulent fragmentation theory, and the scaling relations due to Larson (1981), their work suffers on 2 counts : (i) the stellar masses were calculated by applying the Bondi-Hoyle(BH) accretion theory to collapsing cores. The BH theory as is well-known, assumes that the fluid being accreted originates from a non-self gravitating background. This assumption breaks down in a collapsing core. (ii) The calculation of the velocity with which gas is accreted by the protostar is unclear. Despite this, their model provides a reasonable estimate of the protostellar accretion rate using which a power-law IMF has been derived. In a slightly earlier work (Anathpindika 2011), we had argued on similar lines where accounting for protostellar feedback, and treating the protostellar accretion rate as a free parameter, power-law IMFs were derived starting with different choices of core mass distribution that also included a uniform distribution. We had then sought to underline the importance of physical processes during the protostellar phase and the protostellar multiplicity in determining the final IMF.

The problem of deriving the stellar IMF from first principles is indeed non-trivial and so a semi-analytic approach is useful to ascertain the effects of some crucial parameters. Cooling processes in star-forming clouds most likely hold the key to sub-fragmentation of unstable clumps into prestellar cores. However, owing to computational expenses, simulations about turbulent fragmentation of gas often assume a cold, isothermal gas, which though reasonable to some extent, fails to account for heating due to protostellar feedback and from other possible external sources. Thus regions within a star-forming cloud exhibit substantial variation in gas temperature and therefore, an equally variable dynamical state. Possible effects of variations in gas temperature were neglected by Veltchev \emph{et al.} in their derivation of the core mass function. In the present work these effects are included by generating multiple realisations of clump fragmentation by subjecting them to a number of random choices of gas temperature. Having so derived a distribution of core masses,  the stellar IMF is derived by applying a typical core-to-star conversion efficiency, $\epsilon$, between 5\% - 15\% to each bound core. Observations of star-forming clouds have yielded comparable values for the star-formation efficiency (e.g., Motte \emph{et al.} 1998; Ward-Thompson \emph{et al.}2007). The paper is organised as follows. The process of generating the mass-spectra is described in \S 2, and the spectra are then discussed in \S 3 and \S 4. We conclude in \S 5.

\section[]{Generating the mass functions}
We begin our calculations by assuming a volume of molecular gas composed of the usual cosmic mixture where neutral Hydrogen makes up about 9 parts out of 10, while the remainder is made up by Helium and other elements like Carbon, Nitrogen and Oxygen. The average mass of a gas molecule, $\bar{m}$, is taken to be $4\times 10^{-24}$ gms. The gas within this volume is characterised by its average density, $\rho_{avg}\sim 10^{-21}$ g cm$^{-3}$, temperature, $T_{gas}=30$ K, and the magnitude of turbulent velocity, $v_{0}$. The size, $L$, of the turbulent region was fixed using the velocity dispersion relation due to Larson (1981) so that the total mass within the volume, $M_{tot}\sim \rho_{avg}L^{3}\sim 2.5\times 10^{8}$ M$_{\odot}$. In a second realisation of the problem, the gas was maintained at a lower temperature, $T_{gas}$ = 15 K, and $M_{tot}\sim 6\times 10^{5}$ M$_{\odot}$.  The probability distribution of gas density, $p(y)$, in the turbulent field is
\begin{equation}
p(y)dy = \frac{1}{\sqrt{2\pi\delta^{2}}}\textrm{exp}\Big[\frac{(y - y_{0})^{2}}{2\delta^{2}}\Big],
\end{equation}
where $y_{0}= -\frac{\delta^{2}}{2}$, and $\delta^{2} = \ln(1 + \mathcal{M}^{2}\gamma^{2})$ with $\gamma\sim 0.5$, as suggested by Padoan \& Nordlund (2002). $\mathcal{M}\equiv v_{0}/a_{0}$, is the Mach number of the turbulent field superposed on the gas maintained at a uniform temperature, $T_{gas}$, and sound speed, $a_{0}$. Finally, $y\equiv\ln(\rho/\rho_{avg})$.

The turbulent field is defined by its energy-spectrum, $\mathcal{E}(k)\propto\ k^{-x}$, for $x > 0$, so that the turbulent velocity, $v(k)$, is simply
\begin{equation} 
v^{2}(k) = \int_{k} \mathcal{E}(k) dk \equiv v_{0}^{2}L^{x-1}\sum k^{1-x},
\end{equation}
where the integration and/or summation on the right-hand side extends over the driven wave-modes, $k$. Turbulence was driven over spatial scales between 10 pc and 200 pc, corresponding to wavenumbers between 1 and 8. The length of the fastest growing mode, $\lambda_{fast}$, is given by the following equation,
\begin{equation}
\lambda_{fast} \sim \Big(\frac{\pi a_{eff}^{2}}{G\bar{\rho}}\Big)^{1/2};
\end{equation}
$\bar{\rho}\sim \mathcal{M}^{2}\rho_{avg}$, where turbulence-induced shocks are assumed to be isothermal. The effective sound-speed, $a_{eff} = (a_{0} + v_{k})$, so that a slight manipulation of Eqn. (3) after substituting for $\bar{\rho}$ and $a_{eff}$, leads us to 
\begin{displaymath}
\lambda_{fast} = \frac{\lambda_{J}}{\mathcal{M}}\Big(1 + \Big(\frac{k\mathcal{E}(k)}{a_{0}^{2}}\Big)^{1/2}\Big),
\end{displaymath}
where $\lambda_{J}$ is the thermal Jeans length for sound-speed, $a_{0}$. For a highly supersonic gas, the left-hand side of the above expression may be approximated to
\begin{equation}
\lambda_{fast}\sim \frac{\lambda_{J}}{v_{0}}[k\mathcal{E}(k)]^{1/2},
\end{equation}
or in terms of only the wave-vector as
\begin{equation}
\lambda_{fast}\propto \lambda_{J}k^{(1-x)/2}.
\end{equation}
The scale of fragmentation is set by $\lambda_{fast}$, so that the mass of a clump, $M_{clump}$, condensing out is of the order,
\begin{equation}
M_{clump}\sim \lambda_{fast}^{3}\bar{\rho},
\end{equation} 
or 
\begin{equation}
M_{clump}\propto k^{3(1-x)/2}.
\end{equation}
The number of clumps, $N_{clumps}$, condensing out is $M_{tot}/M_{clump}$, so that the clump-mass function (MF), is
\begin{equation}
\frac{dN_{clumps}}{dM_{clump}} = -\frac{M_{tot}}{M_{clump}^{2}} \propto k^{-3(1-x)}.
\end{equation}
The MF was generated by Monte-Carlo integrating the above equation with about $10^{5}$ choices of random perturbations in the density field. We note that clump-formation was terminated once $\sum N_{clumps} \gtrsim M_{tot}$, so that mass is conserved. 

The dynamical stability of clumps, it is well-known, is determined by the balance between heating and cooling processes. Possible effects of various cooling processes such as molecular line emission at densities below 10$^{4}$ cm$^{-3}$, and the gas-dust coupling at much higher densities, were introduced in the calculations by subjecting the gas within these clumps to 10$^{3}$ random choices of temperature distributed uniformly between 7 K- 20K, and then 6 K- 15 K, in each of the two realisations. The temperature distribution is characterised by a probability density, $p(T)$, and the core mass function (CMF) was obtained by convolving it with the clump mass function, so that
\begin{equation}
M_{core} = \int_{T} p(T)\frac{dN_{clumps}(\bar{\rho})}{dM_{clumps}} dT.
\end{equation}
Alternatively, $\lambda_{core} = \Big(\frac{a_{0}(p(T))}{a_{eff}}\Big)\lambda_{fast}$, so that $M_{core} \sim \lambda_{core}^{3}\bar{\rho}$.  The number of cores, $N_{cores}$, is then, $M_{clump}/M_{core}$. Finally, the initial mass function, ({\small IMF}), was derived by applying a random core-to-star conversion efficiency, $\epsilon$, between 5\% to 15 \%, to each bound core.

\begin{figure}
 \vspace{10pt}
 \includegraphics[angle=270,width=8.cm]{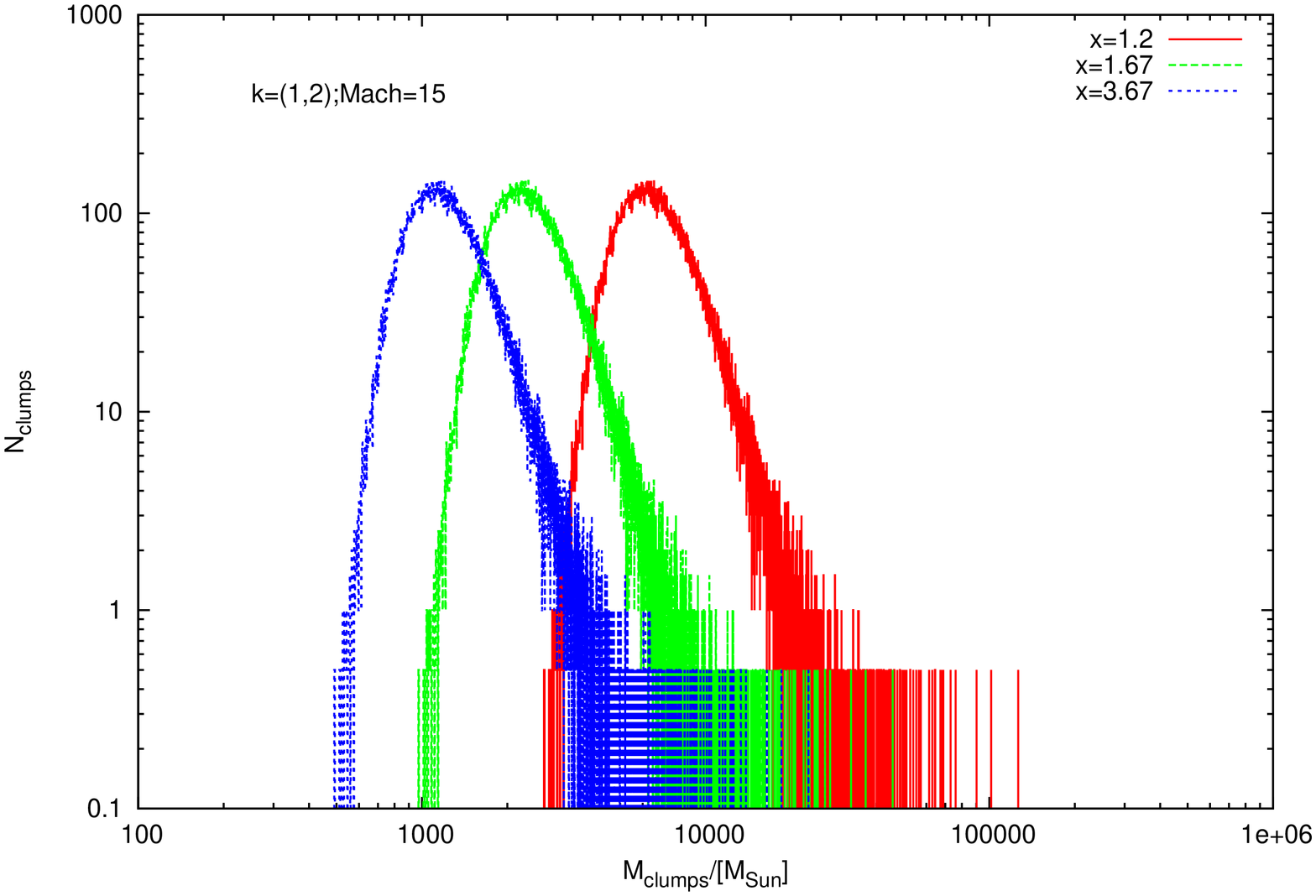}
 \includegraphics[angle=270,width=8.cm]{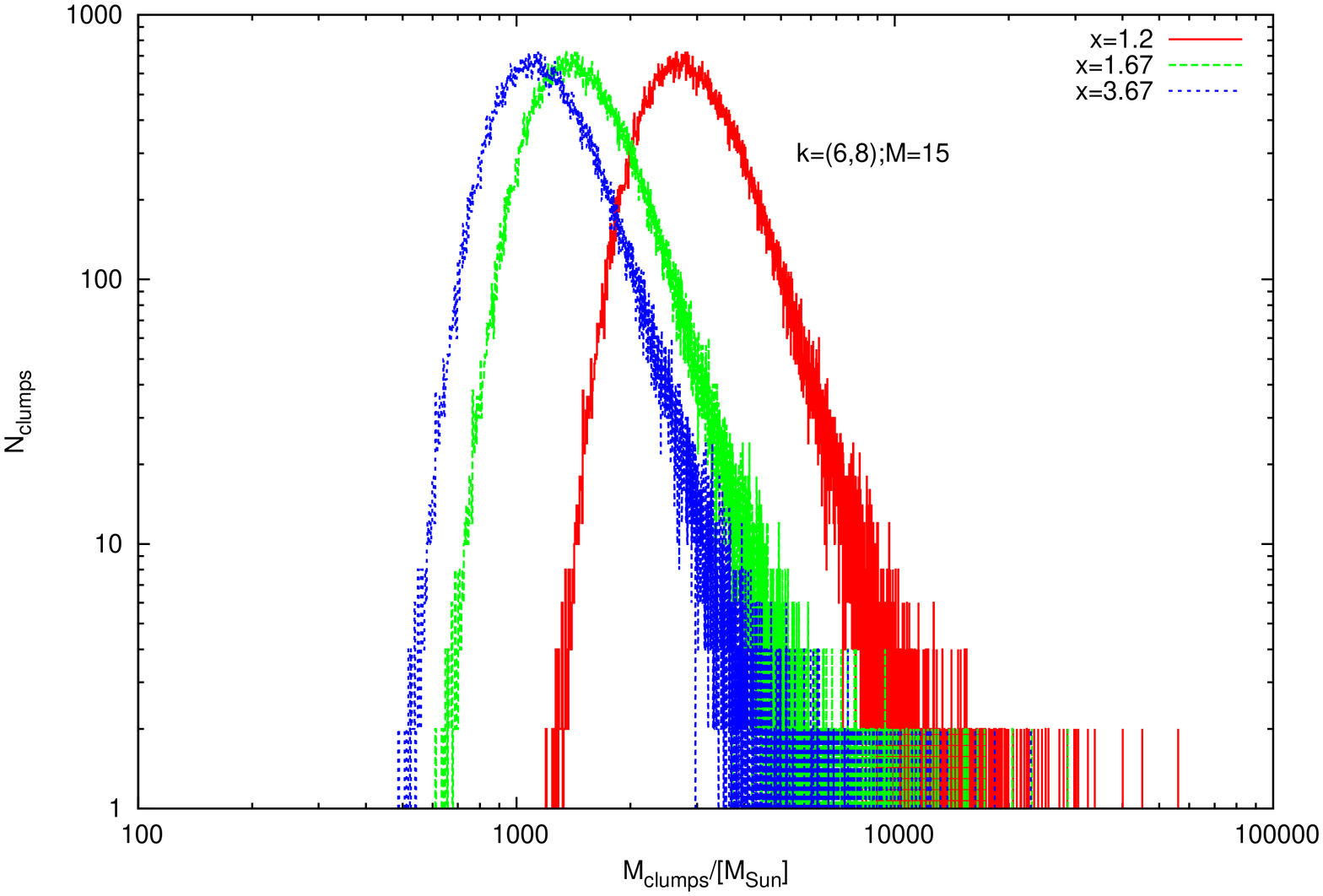}
\caption{Shown in each panel is the derived clump mass-spectrum for different choices of driven wave-modes, $k$, and the index of turbulent power-spectrum, $x$, as marked on each plot. The spectra are roughly lognormal in nature, though with a slight power-law extension at the high-mass end.}
\end{figure}

\section[]{Mass functions (MF)}
Using the basic turbulent fragmentation hypothesis we have derived mass functions (MFs) for dense clumps, cores and stars. Fragmentation of the volume of gas considered here is purely due to the growth of density perturbations that are random in nature. While our interest lies in the clump-formation via fragmentation, the dissipation of turbulent energy is neglected. Results of this exercise are summarised by the plots in Figs. 1 through to 5. The distribution of clump masses derived here has been shown in Fig. 1. Irrespective of the choice of driven wave-modes, the mass-functions exhibit remarkable similarities. The distributions in each case are roughly lognormal, though with a slight power-law extension at the high-mass end. The spectra, however, are significantly steeper than that for typical clumps, which have a slope of $\sim 2$ (e.g. Kramer \emph{et al.} 1998). Also, it follows from Eqn. (5), a relatively shallow energy spectrum ($x$=1.2) shows a greater tendency of producing larger clumps. This can indeed be seen from the plots in Fig. 1

The core mass function(CMF), generated for different choices of the spectral index, $x$, and driven modes, $k$, have been shown in Fig. 2. Overlaid on the plot shown in the upper panel of Fig. 2 is the  power-law CMF defined as,
\begin{equation}
\frac{dN_{cores}}{dM_{cores}} \propto M_{cores}^{-\alpha}
\end{equation}
with,
\begin{displaymath}
\alpha = \left\{ 
  \begin{array}{rl}
   4.2 &; \textrm{if } \frac{M_{core}}{\textrm{M}_{\odot}}\gtrsim 1,  \\ 
   2.1 &;  0.1\lesssim\frac{M_{core}}{\textrm{M}_{\odot}} < 1, \\ 
  -1.9 &;  0.02\lesssim\frac{M_{core}}{\textrm{M}_{\odot}} < 0.1,  \\ 
  \end{array}\right.
\end{displaymath}
The power-law with these exponents fits the derived CMF reasonably well, acceptable at 5\% significance level of the $\chi^{2}$-test, performed with 5 degrees of freedom, viz., the gas temperature, density perturbations, turbulence Mach number, $\mathcal{M}$, spectral-index, $x$, and the wavemodes driven, $k$. 

Evidently the power-law defined by Eqn. (10) is considerably steeper than that for cores found in typical star-forming regions such as the Orion MC (e.g., Nutter \& Ward-Thompson 2007), and the Ophiuchus MC (e.g., Johnstone \emph{et al.} 2001). Increasing the gas temperature and the Mach number, $\mathcal{M}$, raises the fragmentation length and therefore the mass of resulting cores which results in a marginally shallow CMF ($M_{cores}^{-3.5}$, for $M_{cores}/M_{\odot} \gtrsim 1$), as can be seen in the central panel of Fig. 2. Likewise, a shallow CMF also results from driving longer wavemodes, $k=(6,8)$, when comparatively lesser power reaches smaller spatial scales. This latter CMF is shown in the lower panel of Fig. 2. A unique feature of the CMFs derived here is the appearance of a plateau for intermediate masses in the range (0.2,1) M$_{\odot}$, which is probably due to the uniform distribution of temperature used in generating the CMF.

\begin{figure}
 \vspace{2pt}
 \centering
 \includegraphics[angle=270,width=8.cm]{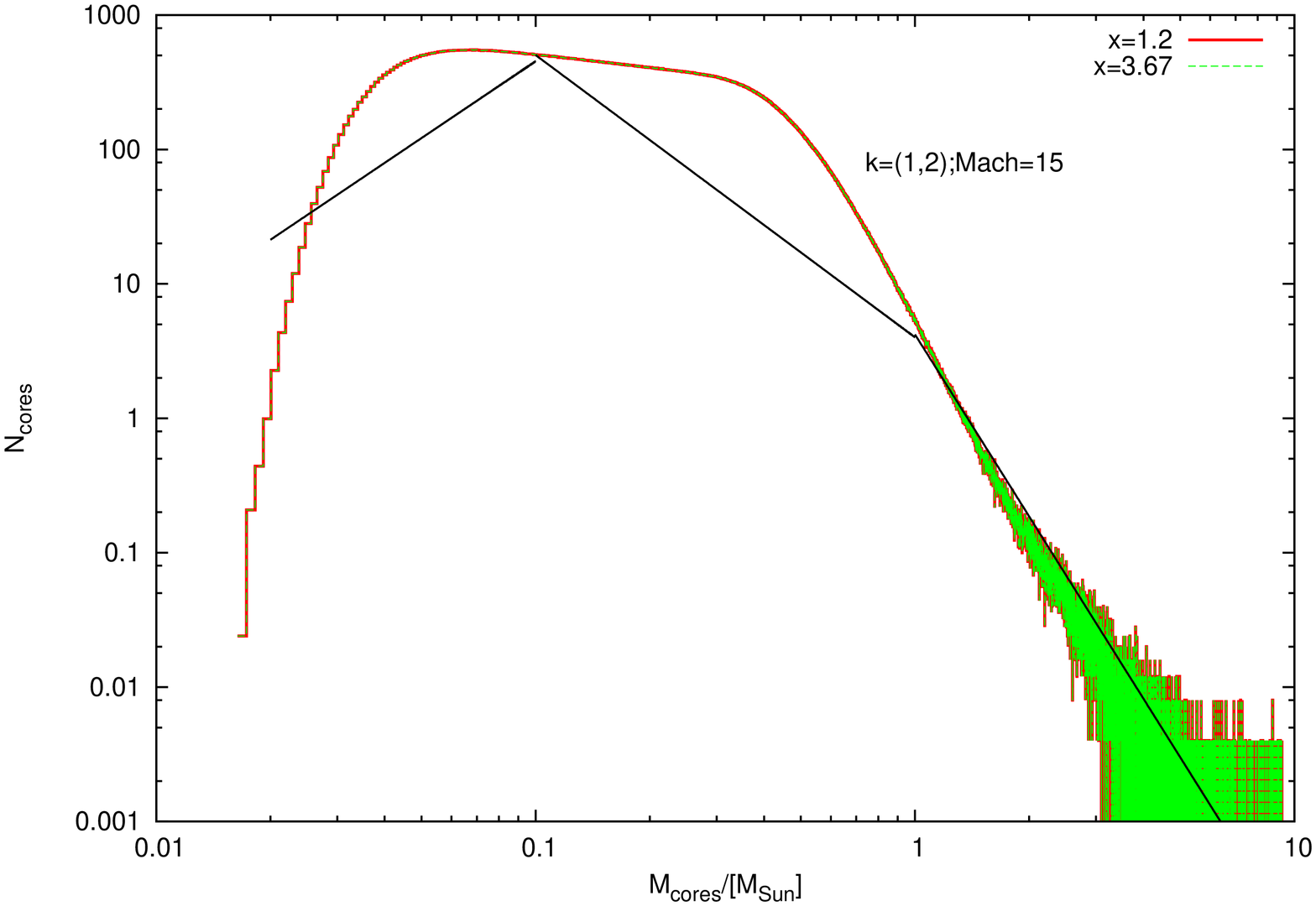}
 \includegraphics[angle=270,width=8.cm]{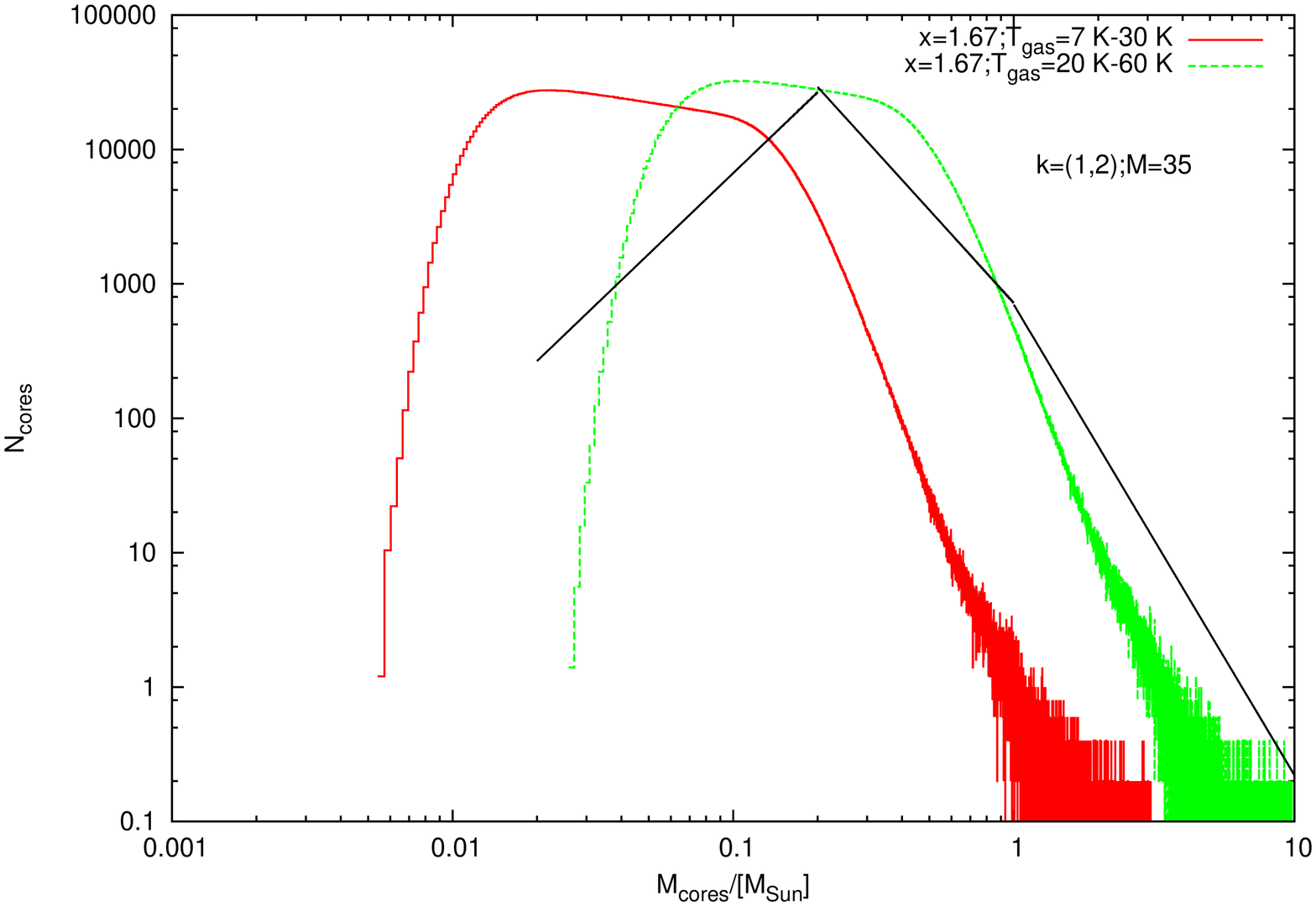}
 \includegraphics[angle=270,width=8.cm]{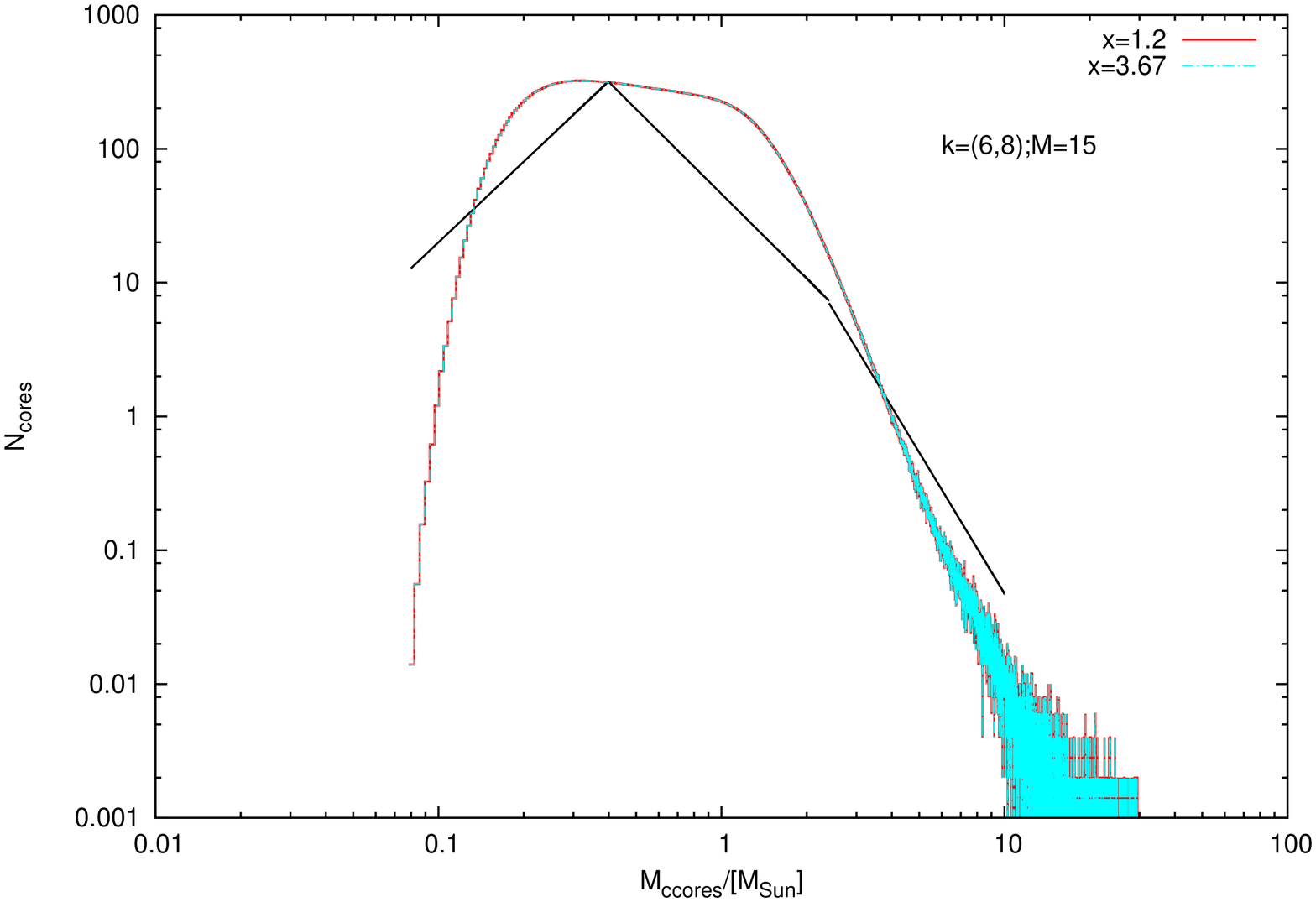}
\caption{Shown in the three panels are the core mass functions derived for different choices of driven modes,$k$, and the power-spectrum index, $x$. Plotted with a continuous black line are the power-laws discussed in the text above. }
\end{figure}

The core-to-star conversion efficiency, $\epsilon$, is the sixth parameter in this scheme apart from the five listed previously. The derived IMF has been shown in the two panels of Fig. 3, where the one in the upper panel was derived from the CMF defined by Eqn. (10). It is significantly steep for $M_{star}/M_{\odot}\gtrsim 0.5$. On the other hand, the IMF derived from the relatively shallow CMF obtained by driving the $k=(6,8)$ modes, has been shown in the lower-panel of Fig. 3. It has been fitted by a power-law defined below, 
\begin{equation}
\alpha = \left\{ 
  \begin{array}{rl} 
    2.7  &;  0.5\lesssim\frac{M_{star}}{\textrm{M}_{\odot}} < 5.0 \\ 
    2.0  &; 0.04\lesssim\frac{M_{star}}{\textrm{M}_{\odot}} < 0.5   \\ 
   -1.0  &; 0.01\lesssim\frac{M_{star}}{\textrm{M}_{\odot}} < 0.04,  \\ 
  \end{array}\right.
\end{equation}
 acceptable at a 10\% significance level of the $\chi^{2}$-test with 6-degrees of freedom. More interestingly, the power-law exponents are within the error-bars on the universal IMF suggested by Kroupa (2002). It can be seen that like the CMF, this IMF also has a plateau in the sub-Solar mass range which most likely reflects a similar feature of the CMF. \\ \\
\textbf{Lowering the initial gas temperature ($T_{gas}$ = 15 K, $\mathcal{M}$=15)} \\
Retaining all other parameters, the mass function for clumps and cores was re-derived in this case, the result of which has been plotted in Figs. 4 and 5. Lowering the initial temperature of the gas, $T_{gas}$, by a factor of 2 to 15 K, does not much alter the overall nature of the two distributions, though massive clumps with mass in excess of $10^{4}$ M$_{\odot}$ now  do not form as can be seen in Fig. 4. As before, gas within these clumps was subjected to a random choice of temperatures in the range 6 K- 15K, and then allowed to fragment. The resulting distribution of core masses has been shown in the upper-panel of Fig. 5. The distribution of core-masses in the lower-panel was derived for a slightly higher initial Mach number ($\mathcal{M}$ = 20). Not only is the CMF in this latter case relatively shallow ($\propto M_{cores}^{-3.5}; M_{cores}\gtrsim 1$ M$_{\odot}$), but the plateau at intermediate masses is also less wide.

\begin{figure}
\centering
 \vspace{30pt}
 \includegraphics[angle=270,width=8.cm]{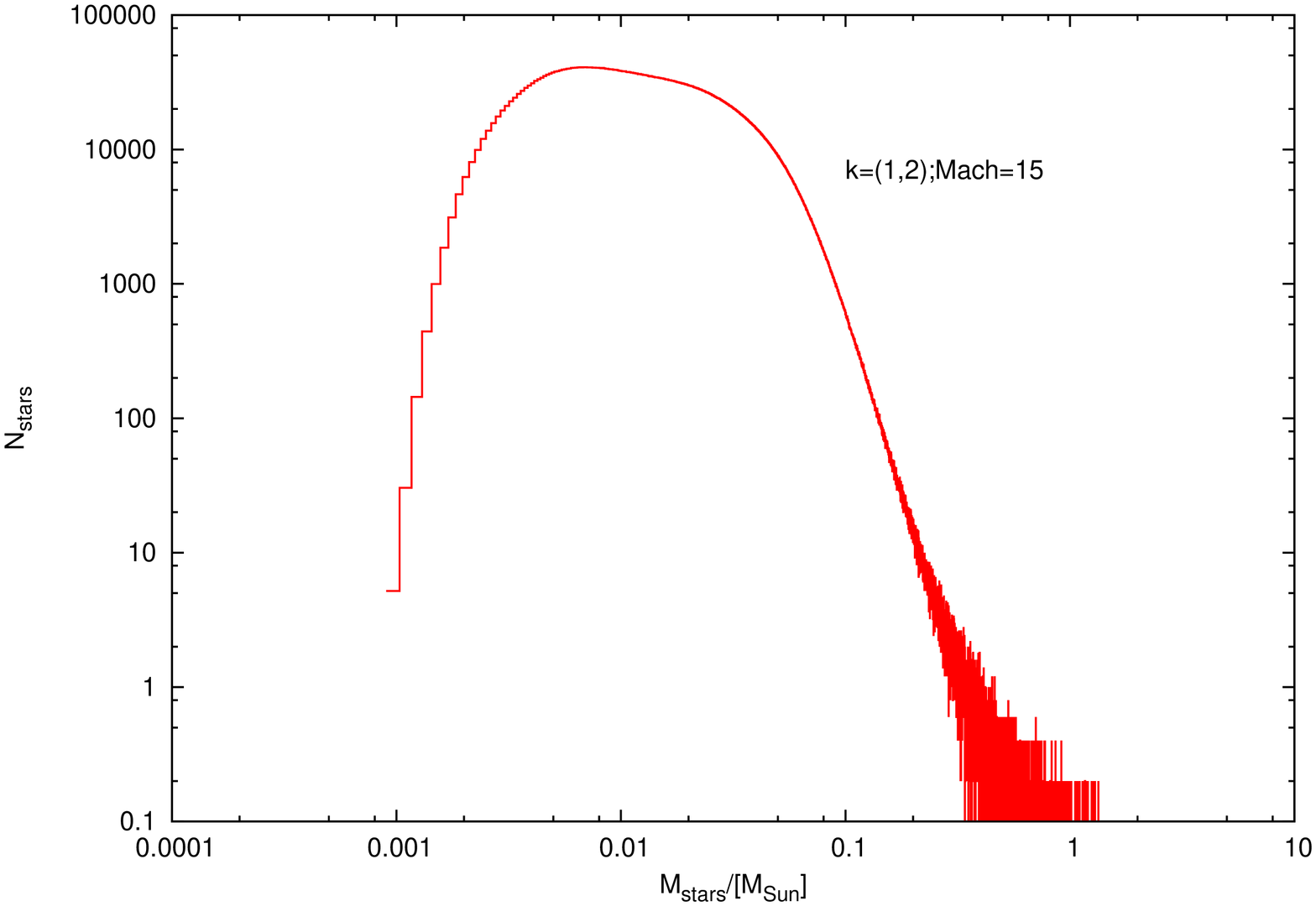}
 \includegraphics[angle=270,width=8.cm]{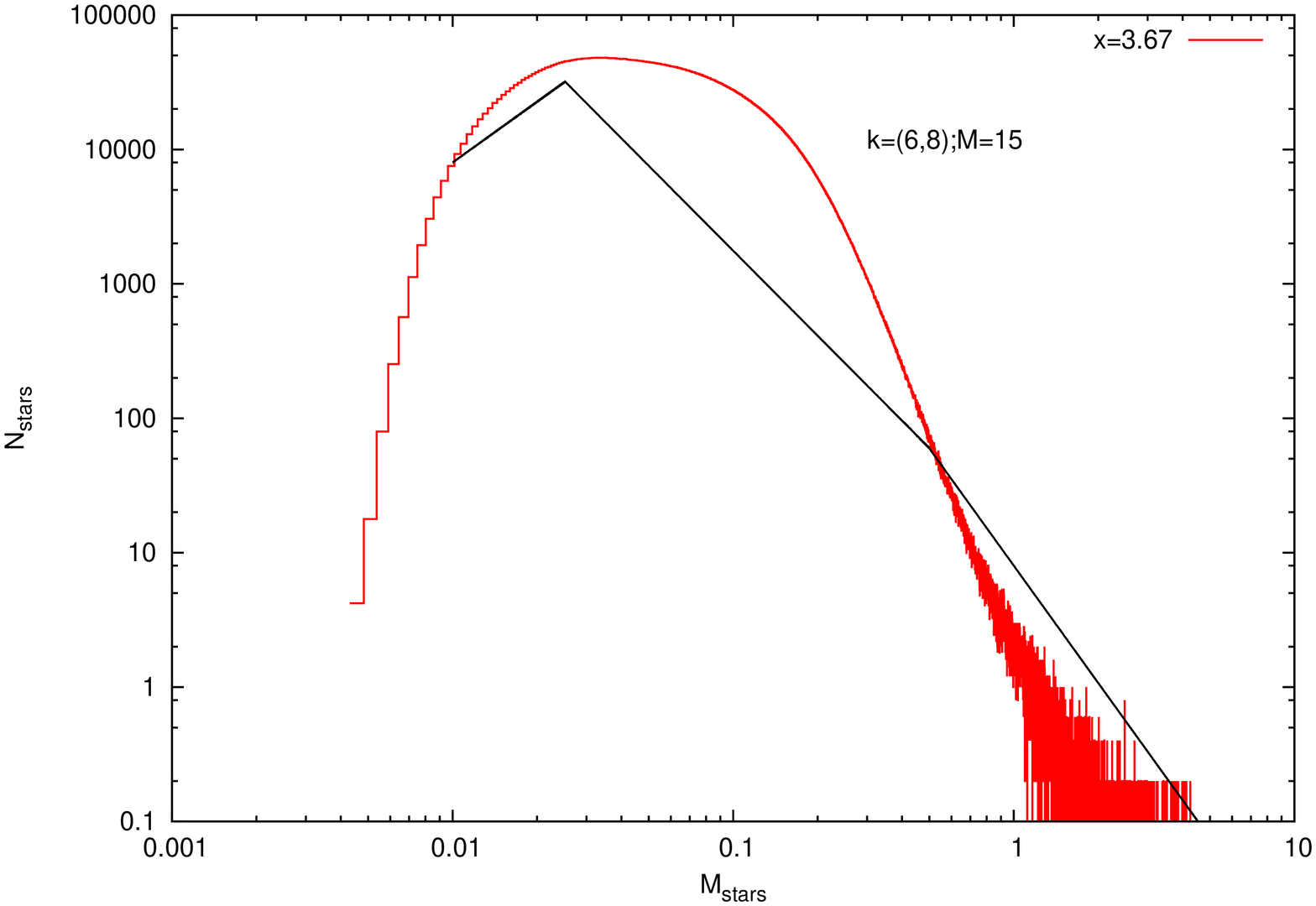}
 \caption{The stellar initial mass function (IMF) derived for two choices of $k$ has been shown in the two panels here. The power-law fit to the derived spectrum in the lower panel is defined by Eqn. (11) in the main text.       }
\end{figure}

%--------------------------------------------------------

\begin{figure}
 \vspace{2pt}
 \centering
 \includegraphics[angle=270,width=8.cm]{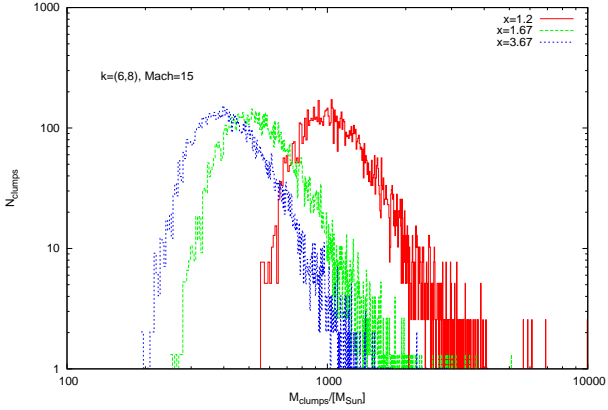}
 \caption{The clump-mass spectrum derived in the next case where the gas was originally maintained at $T_{gas}$= 15 K. Apart from a broader full-width at half-maximum, these spectra do not differ much from those plotted in the upper panel.} \nonumber
\end{figure}

\section{Discussion}
In the previous section, assuming a set of demonstrative parameters we derived mass distributions for clumps, prestellar cores and stars by invoking the theory of turbulent fragmentation. Analogous to the PN02 model, we commenced with a lognormal PDF while setting the density threshold as dictated by the jump condition for an isothermal shock. The derived clump mass function largely reflects the density PDF, though it also has a power-law tail at the high-mass end. This tail is relatively shallow in the case with $T_{gas}$ = 15 K. Steepening the power-spectrum index, on the other hand, simply shifts the mass distributions towards lower masses, closer to those of typical clumps.

The core mass function, irrespective of the choice of wavenumbers driven, the temperature of gas within clumps or indeed, the Mach number, has two parts : a Salpeter-like power-law for $M_{core}/\textrm{M}_{\odot}\gtrsim 1$, and a plateau at lower masses. We have also seen that the power-law for  $M_{core}/\textrm{M}_{\odot}\gtrsim 1$ is relatively steep, and lies between 3.5-4 which is probably the result of an analogously steep mass-function for clumps which precede the CMFs. Interestingly though, these CMFs extend into the regime of low mass cores ($\lesssim 10^{-1}$ M$_{\odot}$). The plateau observed in the CMFs also appears in the stellar IMFs as can be easily noticed through a comparison between Figs. 2 and 3. The CMFs and the IMFs derived here both show a significant over-population in the (0.5,1.0) M$_{\odot}$ range. The plateau in the CMFs derived here merely reflect the distribution, $p(T)$, of gas temperature within clumps which, we remind, was assumed to be distributed uniformly between 7 K - 30 K, and then between 6 K- 15 K in the next realisation. The plateau can be avoided by confining the gas temperature to a fairly narrow range implying that, the temperature of gas across star-forming clumps is unlikely to show much variation. Of course, we are discounting regions which may exhibit extreme conditions.

Fragmentation in a volume of turbulent gas proceeds along a hierarchy that begins with clumps and filamentary clouds. Some of which fragment further to produce prestellar cores. Indeed, some recent maps of star-forming regions in various sub-millimetre wavebands have revealed extensive networks of filamentary clouds with a number of prestellar cores (e.g. Andr{\' e} \emph{et al.} 2010; Men'shchikov \emph{et al.} 2010). From the preceding analysis in \S 2 it is clear that the prevalent physical conditions within gas clumps are crucial in determining the proclivity towards further fragmentation. To the first order, the minimum mass of fragmentation is set by the thermal Jeans mass, $M_{J}\sim \bar{\rho}\lambda_{J}^{3}$. In reality of course, the magnetic field will revise this mass upward, however, in the present exercise we do not include the effects of magnetic field.

Gas temperature and density are locked in a reciprocally causative relationship. It is in this respect that a density threshold assumes critical significance, for cooling mechanisms are density dependent. Atomic cooling for instance, dominates in the relatively low-density gas up-to a few hundred particles per cubic centimetre, then the coupling between gas and dust dominates up-to densities of a few thousand particles per cubic centimetre and at even higher densities such as those in typical cores ($\gtrsim 10^{5}$ cm$^{-3}$), molecular species freeze out. Similarly, effects of magnetic flux diffusion are likely to become significant at densities upward of several thousand particles per cubic centimetre. Taken cognitively, these effects can raise the minimum mass of fragmentation considerably (e.g. Larson 2005). 

\begin{figure}
  \includegraphics[angle=270,width=8.cm]{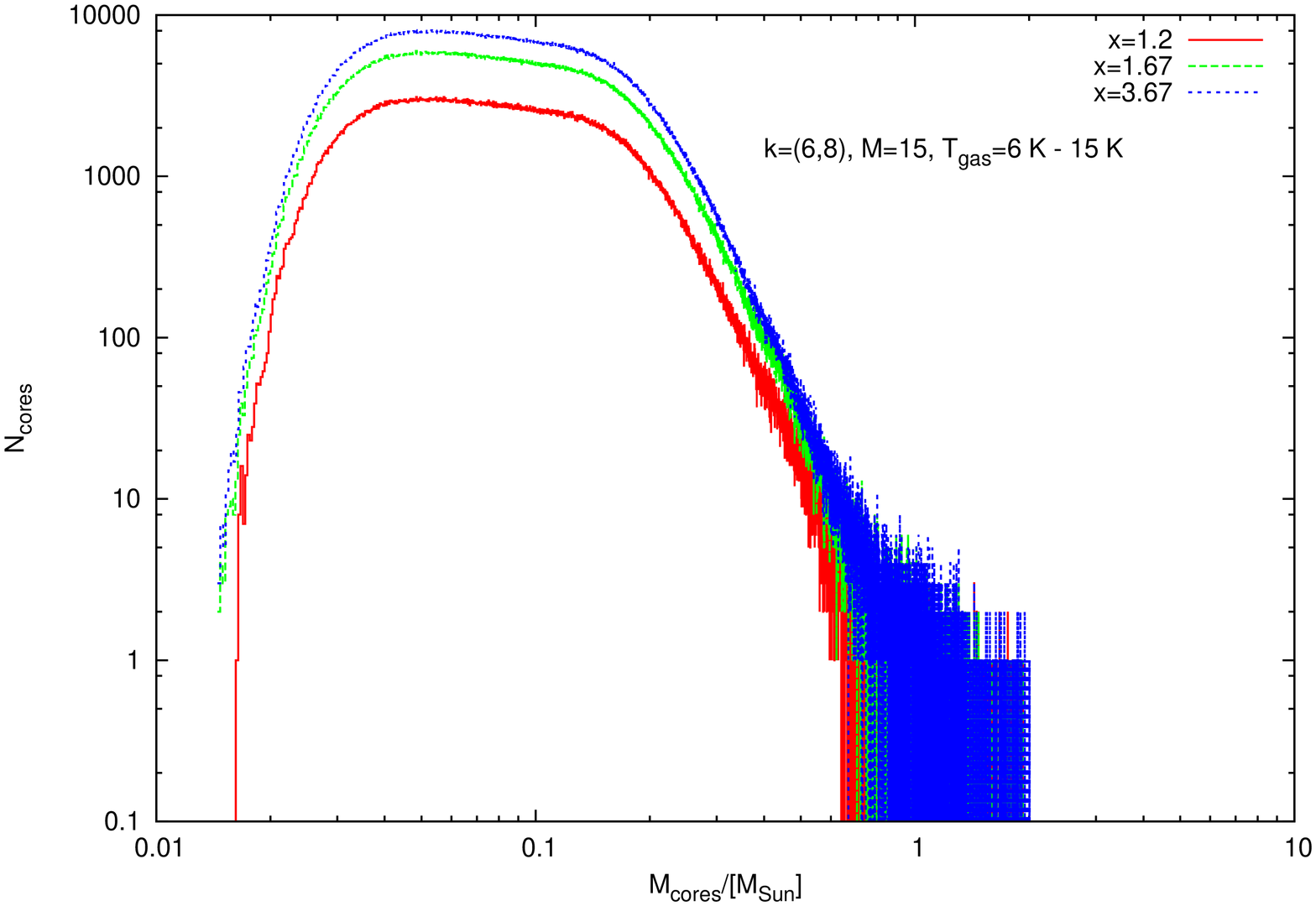}
   \includegraphics[angle=270,width=8.cm]{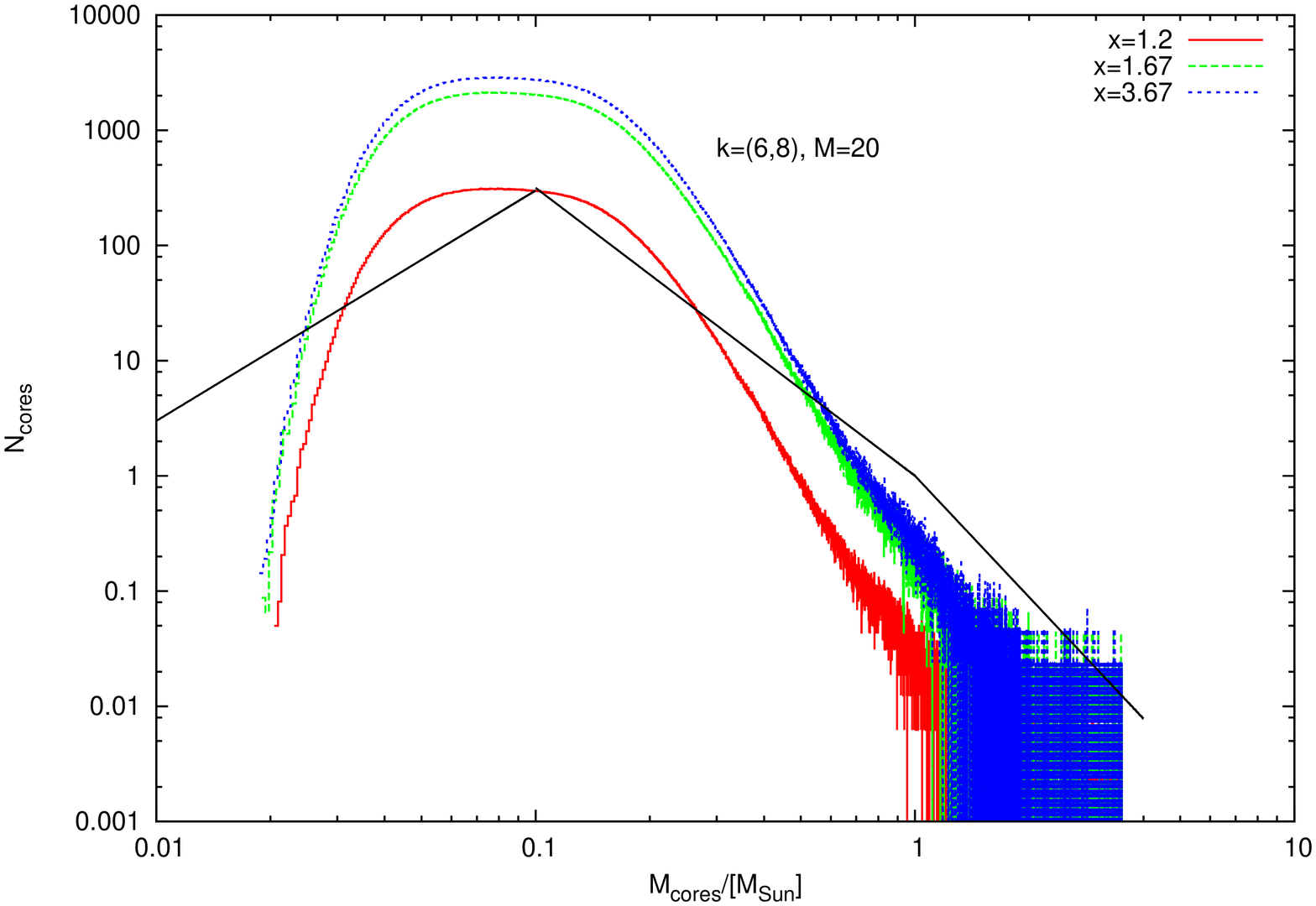}
 \caption{Shown in the upper and lower-panel of this figure are the CMFs derived in the second realisation of the problem where the original gas was maintained at 15 K. A power-law fitted to the derived distribution has been plotted on top with a black line.}
\end{figure}

Here we have only included the effects of gas thermodynamics by accounting for the gas temperature and the velocity dispersion. The contribution of the velocity dispersion to the fragmentation length scale is determined by the energy-spectrum, $\mathcal{E}(k)$. In general, the impact of the energy-spectrum is readily visible on the mass-function for clumps where relatively smaller clumps form with steepening energy-spectrum; see plots in Fig. 1. The CMFs on the other hand, while exhibiting immunity to changes in $\mathcal{E}(k)$, are significantly affected by changes in the thermal state of the gas, as demonstrated by plots shown in the central panel of Fig. 2. This is further corroborated by results from the exercise where the temperature of the pre-fragmentation gas was lowered by a factor of 2; see also Figs. 4 and 5. In either case, temperature of the gas within a clump was treated as a random variable to study the effects of its variation. The gas temperature thus, was allowed to take values  within a range typically found in star-forming clouds such as Taurus (e.g., Falgarone \emph{et al.} 1998), Ophiuchus (e.g., Johnstone \emph{et al.} 2000), and Orion (e.g., Yu \emph{et al.} 1997). 

 Prestellar cores collapse under favourable conditions to form stars and dwarfs, though the collapse phase itself is associated with competing processes such as feedback from young stellar objects, competition between accreting stars to acquire mass and the interaction between a young star and its attendant disk. These processes, though highly non-linear in nature, are critical in determining the stellar IMF. In literature, these processes have sometimes been grouped together as the core fragmentation mass function (e.g. Elmegreen 2000; 2011). Clearly, stellar multiplicity is an essential component of the stellar IMF (e.g., Goodwin \& Kouwenhowen 2009). In an earlier work (Anathpindika 2011), we had presented a model of core fragmentation that included the details listed above, and therefore the prospect of forming brown-dwarfs via core-fragmentation. In view of the non-triviality about estimating the mass of a protostellar disk, that model failed to predict any objects possibly resulting from disk-fragmentation. However, assuming the rate of protostellar accretion as a free parameter, we had then demonstrated that the processes during the collapse of a core hold the key to determining the nature of the IMF. That model had also demonstrated the robustness of the IMF to changes in the preceding CMF. In the present work though, we have derived the IMF by applying a typical core-to-star conversion efficiency, $\epsilon$, to each bound core. In the present case it is therefore not possible to comment on the companion stars. However, it is clear that the power-law tail in the CMF reappears in the IMF, albeit with a slightly shallower slope and with significant changes to the plateau at sub-Solar masses. 

The number of parameters used in the current exercise could invite some criticism, but these parameters are essential components of the turbulent fragmentation model and adopted from recent reports about star-forming clouds. For a typical choice of parameter values, the derived mass functions are broadly similar to those determined observationally. The agreement was best seen for results obtained with - (i) a steep power-spectrum, typically steeper than the Kolmogorov-type, up to about $k^{-3.67}$, and (ii) by driving longer wavemodes, $k=(6,8)$. In view of (i), we suggest that the power-spectrum index for gas in MCs could indeed be steeper than Kolmogorov-type. Second, turbulence is likely to be driven on larger scales which is consistent with observations and has also demonstrated by Brunt \emph{et al.} (2009) through hydrodynamic simulations. Furthermore, studies of the ${}^{13}$CO emission line for the Cygnus star-forming region (Schneider \emph{et al.}2011), and a number of other star-forming clouds in the Galactic ring survey (Roman-Duval \emph{et al.} 2011), have shown that turbulence is indeed driven on larger spatial scales, typically on the scale of the the cloud size. A point that has not been addressed in the present work is about the likely effects of changes in the gas density. However, from Eqns. (3) - (6) it is quite clear that lower density may very well produce predominantly larger clumps, which might possibly explain the relatively steep CMFs observed in low-density clouds. It might also explain the steep stellar mass distributions in some spiral galaxies with peculiar properties, such as those with low surface brightness (e.g. Lee \emph{et al.} 2004). Thus, if indeed the stellar IMF has a universal nature then the underlying processes which generate it could also be similar. Gravoturbulent fragmentation, it appears, could play a key role in assembling the initial conditions for star-formation. This theory also admits the possibility of rapid star-formation which makes it more attractive.

\section{Conclusions}
Using the concept of turbulent fragmentation we have derived the mass spectra for clumps, prestellar cores and stars via a Monte Carlo integration of the underlying probability distribution. We first calculated the clump mass spectrum by allowing the random density perturbations to condense out as clumps, while the CMF was derived through the convolution of this mass spectrum with a random distribution of gas temperature for the dense clumps. Irrespective of the initial gas temperature, we observed that best agreement with observational findings resulted from driving longer wavemodes and with a power-spectrum steeper than Kolmogorov-type. Thus, while fragmentation of the original volume of gas produces clumps which themselves may break-up further, chemical processes that determine gas temperature in the dense phase are likely to play a crucial part, leading up to the distribution of core masses. In fact, we also suggest that the characteristic knee seen in the CMF and the stellar IMF could be a pointer to the temperature of the gas within star-forming clumps, and therefore, the efficiency of cooling processes. For the theory of fragmentation to predict the core masses more accurately, particularly at the low-mass end, it must be coupled with appropriate thermodynamic details of the density field. Arguably, this is a non-trivial task, and so, in the present work temperature has been substituted as a random variable that takes values similar to those found in typical star-forming clouds.

 The IMF was also derived in a similar fashion by treating the star-formation efficiency as a random variable. The IMF so derived was found to be in reasonably good agreement with the universal IMF, acceptable at a 10\% significance level of the $\chi^{2}$-test with 6 degrees of freedom, viz.  the gas temperature, density perturbations, the Mach number and spectral-index of the turbulent field, the wavemodes driven, and the core-to-star conversion efficiency. It is therefore clear that apart from the physical properties of the gas, the history of protostellar collapse is critical towards determining the nature of the IMF. Our work shows that a typical CMF, and therefore an IMF, emerge quite naturally from the application of the theory of turbulent fragmentation by driving longer wavemodes. This claim is corroborated by numerical and observational results presented by Brunt \emph{et al.} (2009) and Schneider \emph{et al.} (2011). While isolated regions within star-forming clouds may exhibit extreme physical conditions leading to corresponding variations in the IMF, results derived here suggest that most star-formation in the nearby clouds is unlikely to vary much; see also the review by Bastian \emph{et al.}(2010). 

\section*{Acknowledgements}
The author acknowledges support via a post doctoral fellowship of the Department of Science \& Technology, Government of India, held at the Indian Institute of Astrophysics. Useful suggestions from an anonymous referee towards the improvement of the original manuscript are gratefully acknowledged.

\label{lastpage}
\end{document}